# Numerical modeling of material points evolution in a system with gravity

Melkikh A.V.*, Melkikh E.A., Kozhevnikov V.A.

Ural Federal University, 620002, Mira str. 19, Yekaterinburg, Russia,

*melkikh2008@rambler.ru

The evolution of material points interacting via gravitational force in 3D space was investigated. At initial moment points with masses of 2.48 Sun masses are randomly distributed inside a cube with an edge of 5 light-years. The modeling was conducted at different initial distributions of velocities and different ratios between potential and kinetic energy of the points. As a result of modeling the time dependence of velocity distribution function of points was obtained. Dependence of particles fraction which had evaporated from initial cluster on time for different initial conditions is obtained. In particular, it was obtained that the fraction of evaporated particles varies between 0,45 and 0,63.

Mutual diffusion of two classes of particles at different initial conditions in the case when at initial moment of time both classes of particles occupy equal parts of cube was investigated.

The maximum Lyapunov exponent of the system with different initial conditions was calculated. The obtained value weakly depends on the ratio between initial kinetic and potential energies and amounts approximately $10^{-5}$. Corresponding time of the particle trajectories divergence turned out to be 40-50 thousand years.

# Introduction

Studying of dynamics of non-stationary gravitating systems is an actual problem for today. Such systems include, for example, stars, star clusters, galaxies and other massive objects.

Mutual influence of bodies interacting on the basis of gravitational field leads to the chaotization of their orbits. Between gravitating bodies correlations exist, which is caused by the random convergence of bodies as a consequence of which the significant change of their trajectories can occur. As a result of substantial convergence, bodies (material points) can gain enough energy to leave the system.

The behavior in time of a large number of bodies interacting through gravitational potential can be described by distribution functions. The problem is that the evolution of the distribution function for the system of bodies with gravity depends significantly on the initial conditions. At the moment, this relationship has not been studied enough.

Thus, the evolution of the distribution function of material points in three-dimensional space is an urgent task. The aims of this article are

- On the basis of numerical simulations to investigate the evolution of the system at different initial conditions (different initial ratios of kinetic and potential energies of the system, the initial distribution of the coordinates and velocities) and to obtain the time dependence of the particle distribution function and the time dependence of the fraction of evaporated particles from different initial conditions
- to investigate the mechanisms of mixing of classes of particles in the system at different initial conditions;
- assess the degree of chaos as a result of the evolution of the system through the maximum Lyapunov exponent

# 1. Dynamics of the systems with gravity

Systems with gravity relate to systems, which are based on long-range potential. The additivity of internal energy of the system is not typical for these systems, because the interaction with distant parts of the system cannot be neglected. As a result, such systems may stay in the long-lived quasi-stationary states, and the term "local equilibrium" for these systems cannot be introduced. Quasi-stationary states are reached in the process of collisionless relaxation.

One of the main postulates of thermodynamics states that the entropy and energy should be additive with respect to subsystems - that is, the interfacial contribution should be negligible. For systems with short-range forces this condition is fulfilled - in the thermodynamic limit interfacial energy is much less than the energy of the bulk (volume) of the substance. This, however, does not refer to systems with long-range forces, for which the interfacial volume cannot be clearly defined - every particle interacts with every other particle system, so that there is no clear separation of the volume and the border (see, eg, [1]).

In the absence of correlations, the dynamic evolution of the distribution function of the system with long-range interactions is determined by collisionless equation of Boltzmann-Vlasov. One of its most important features is that the system described in this equation does not come into a state of thermodynamic equilibrium (in contrast with systems with short-range potentials).

As a model systems for numerical simulation clusters of galaxies, galaxies, star clusters, asteroids and other objects are used [2-9].

Systems with long-range potential were reviewed in [1, 11]. In particular, the systems with gravity for one-, two- and three-dimensional space were considered. Numerical and mathematical modeling showed that the system with gravity does not come to a state of equilibrium, and can exist for a long time in one of the quasi-stationary states. However, the results only relate to a system with spherically symmetric initial conditions. Wherein transition evolution of quasi-stationary states remains largely unclear.

In the works [4,5] the Lyapunov time and decay time of the system of three bodies, as well as the dynamics of weakly hierarchical triple stars of equal masses were investigated. For the initial conditions near the resonance (initial ratio of the orbital periods of outer star and inner binary are close to 2 : 1) and away from it the dependencies Lyapunov time - the decay time of the system are constructed. It is shown that that near resonance and away from it the nature of these relationships varies considerably.

The evolution of star clusters was investigated in [12]. The hypothesis of the formation of some moving clusters of stars which left as a result of the collapse of the star clusters, was verified on the Hyades cluster as an example. As a result of numerical experiments conducted, it was revealed that in the vicinity of the Sun the stars, which previously belonged to the Hyades cluster, and in the process of dynamic evolution left the cluster, can be observed. The number of such stars does not exceed a dozen.

Thus, the numerical simulation of the evolution of the distribution function of the material points in the self-gravitating systems remains largely unclear and remains an urgent task.

## 2. Setting of the problem of the system with gravity modeling

Let us consider a system of bodies interacting according to the laws of classical gravity, and investigate the evolution of the distribution function.

All bodies are treated as material points, which interact with each other only through the gravitational forces in space, and the equations of motion are calculated according to the Newton's second law, where the force - is the gravitational attraction from the surrounding bodies:

$$m_i \frac{d^2 \vec{r}_i}{dt^2} = -G \sum_{j=1}^{N} \frac{m_i m_j}{\left(\vec{r}_i - \vec{r}_j\right)^2} \frac{\vec{r}_i - \vec{r}_j}{\left|\vec{r}_i - \vec{r}_j\right|} \qquad (1.1)$$

where $m_i$ – the mass of *i*-th particle, $\vec{r}_i$ – radius vector of *i*-th material point, $N$ – number of particles.

The gravitational potential and kinetic energy of the whole system are the following, respectively (see, e.g., [15]):

$$U = -G \sum_{i=2}^{N} \sum_{j=1}^{i-1} \frac{m_i m_j}{r^{(ij)}}, \qquad (1.2)$$

$$K = \sum_{i=1}^{N} \frac{m_i \vec{v}_i^{\,2}}{2}, \qquad (1.3)$$

where i, j - indexes of bodies, $r^{(ij)}$ - distance between i-th and j-th material point, $\vec{v}_i$ - velocity vector.

Because, as it has been shown previously, the initial conditions play in the evolution of such systems a special role, it makes sense to consider the evolution of the distribution function at significantly different initial conditions.

In this paper, to study the evolution of systems with gravity such initial conditions as «waterbag» with massive particles (m = 2,48 masses of the Sun - MS), accepted as material points have been used (see, eg, [1], [10]). This model represents a condition in which all particles are in a some small limited region of space. The system consists of 400 particles distributed in a cube with the linear dimensions of the 5 light -years. The velocities of particles are equal and directed randomly. Table 1 lists the types of initial conditions used.

Table 1

|   | The ratio of kinetic and potential energy | The initial distribution of the particles in space |
|---|---|---|
| 1 | $2K = -U$ | Uniform random distribution |
| 2 | $8K = -U$ | Uniform random distribution |
| 3 | $K = -5U$ | Uniform random distribution |
| 4 | $2K = -U$ | Location of particles in a regular grid |

The calculations were based on the velocity form of the Verlet algorithm (see, eg, [13, 14]). The peculiarity of the calculations for systems with gravity is that there are no collisions between particles, but their close encounters are possible. Exactly as a result of such encounters particle velocity changes most strongly. To ensure the correctness of the calculations, the step of calculation was chosen such that the error caused by close encounters was small compared with the error caused by the random setting of initial conditions.

The algorithm of calculations can be represented as follows:

1) In the first step the initial coordinates $\vec{r}_i$ and $\vec{v}_i$ velocities for finite number of particles are generated (i = 1, .., N), the masses of particles are set. All information about the particles is saved in the data array. A value of the integration step and the time variable is set to $t_k = 0$.

2) The force acting on the particle is considered as the sum of the attraction forces of other bodies (according to the law of gravity)

$$\vec{F}_i = \sum_{i \neq j}^{N} \vec{F}_j \quad (1.4)$$

where $F$ is an attracting force, $N$ – the number of particles in the system.

3) All the found forces for each particle are stored in an array of data for subsequent calculations.

4) The solutions of the equations of motion for each particle on the time step $t_{k+1}$, are performed, new coordinates and velocities of the particles are calculated:

$$\frac{d\vec{r}_i}{dt} = \vec{v}_i, \qquad (1.5)$$

$$\frac{d\vec{v}_i}{dt} = G \sum_{j \neq i}^{N} m_j \frac{\vec{r}_j - \vec{r}_i}{\left|\vec{r}_i - \vec{r}_j\right|^3}, \qquad (1.6)$$

where , $\vec{v}_i$ - the radius vector and velocity of the i-th body, respectively, $\vec{r}_j, m_j$ - the radius vector and the mass of the j-th body, $G$ - the gravitational constant.

5) Saving of new coordinates and velocities of the particles which were found on the fourth step in the array of data.

6) The value of the variable for time is incremented $t_k = t_{k+1}$, the process moves to step (2) if $t_k < T$ (T - the final simulation time).

# 3. Numerical simulation of the evolution of gravitating systems with different initial conditions

Before the experiments it was necessary to establish a simulation step through the convergence condition so that the result of the error did not exceed 1% over the maximum time of calculation. An experiment was performed with a different time step and the value was obtained at which the specified accuracy is achieved - 0.0002 years. Step simulation is selected fixed.

## 3.1. A random initial distribution of particle coordinates with the ratio of the kinetic and potential energy of 2K = -U

To study the evolution of gravitational system a random initial coordinates distribution of the particles in a 3D cube-shaped cluster was considered. It has linear dimensions of 5 light years. The kinetic and potential energy of the system include both *2K = -U* (condition of the Virial theorem).

As we know, the evolution of an isolated system of bodies in which the total energy is conserved is associated with the fulfillment of the virial theorem:

$$2K + U = 0 ,$$

where *K* and *U* - total kinetic and potential energies of the system.

Virial theorem expresses a balance between the tendency of the gravitational force to compress the system and tendency for expansion because of the velocity of the stars. At high kinetic energies set of material points represents almost ideal gas, whose particles easily fly into space. In the opposite case, a fairly tight cluster is formed, which particles can leave as a result of energy exchange.

The components of the velocity of the particles are equal in magnitude and equiprobable in direction. Velocities module of all particles is 10 km / s.

Figure 3.1 shows the initial distribution of particles in the cluster for the cases with primary energy ratio *2K = -U, 8K = -U, 5U = -K.*

As a result of the evolution distribution function of the velocity and coordinates changes with time. From simulation results velocity distribution function for 3 time marks - 25, 50, 200 thousand years – is constructed.

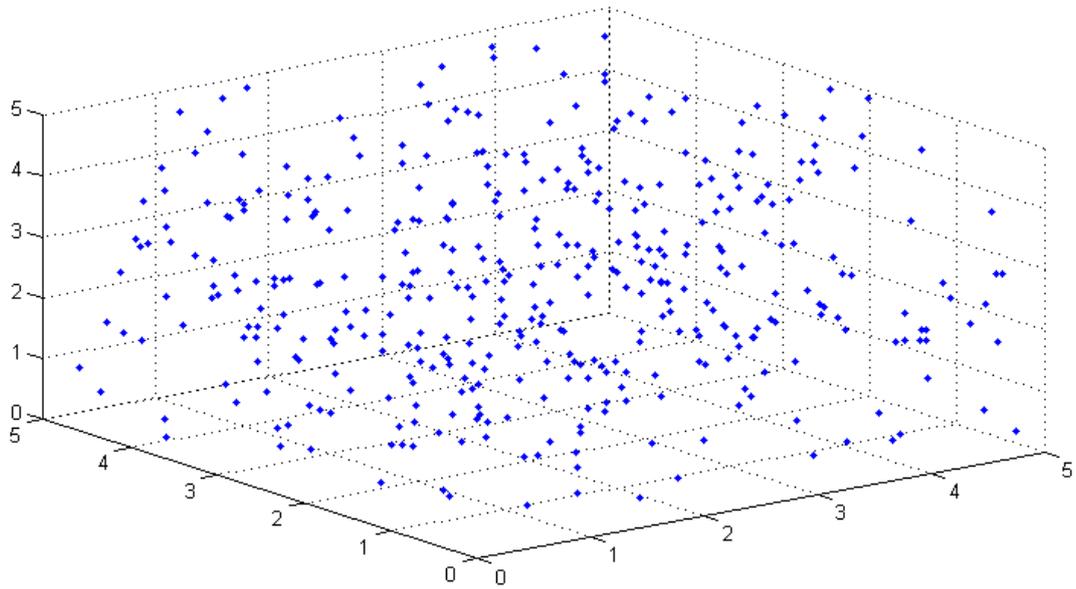

Fig. 3.1 – The random distribution of the particles in 3D cluster at the initial time.

Figure 3.2 shows that the velocity distribution function changes with time its form. Since the initial particle distribution was random and dense enough, in the initial stage of evolution of the system we can observe nucleus formation - the greater part of the particles directs to the center of mass of the system (Figure 3.3).

It is because of this, all particles acquire a much higher velocities than the initial ones, which can be seen in Figure 3.2 at the time of the simulation time 25 thousand years.

As can be seen from Fig. 3.2 the velocity distribution in the course of evolution shifts toward lower velocities.

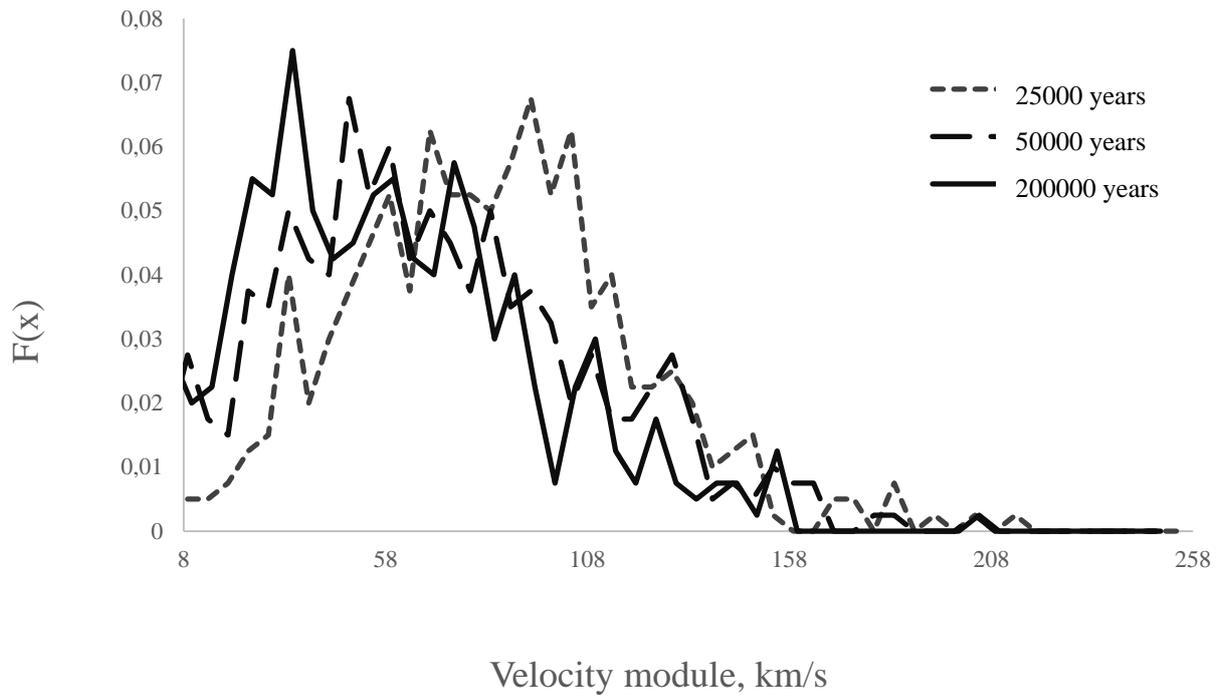

Fig. 3.2 - The distribution of material points on the velocities in the process of evolution. F(x) is the density of probability, normalized value.

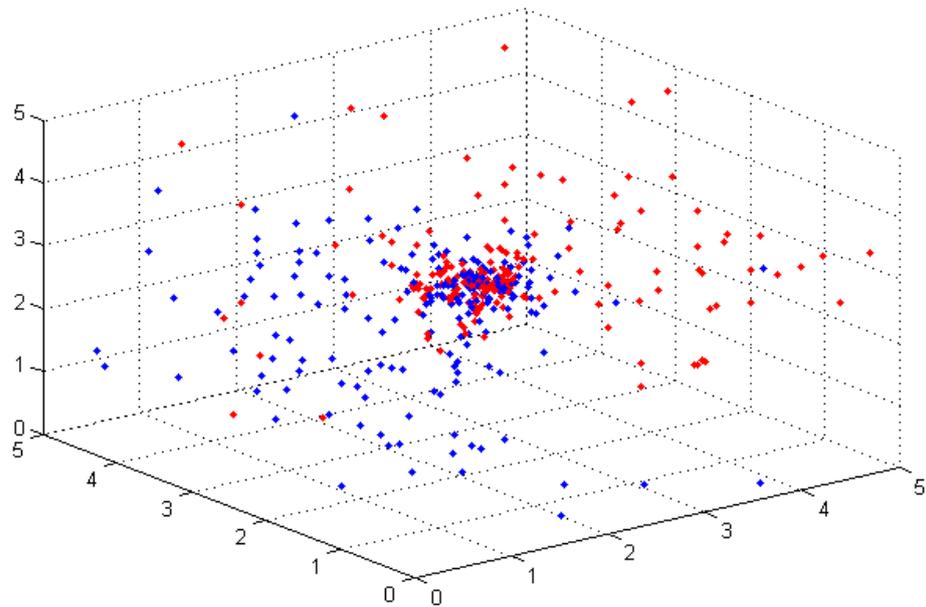

Fig. 3.3 – The distribution of particles in the cluster beyond 25 thousands of years after the start of evolution

Such behavior of stars takes place in the real star clusters. Wherein, in the process of the evolution the stellar pairs are formed (see, eg., [16]). Since the gravitational potential at the surface of the cluster is finite, so the stars that have sufficient energy, leave the system (evaporate).

Thus, collisions lead to the evaporation of the stars out of the system. Only a small fraction of stars leaves at high velocity, and the bulk of them slowly spreads out and forms a tenuous halo star cluster. At the same time the remaining mass of stars lose energy, compress, forming the condensing core. Isolated star system (by taking into account only binary collisions) should finish its existence by evaporation of most of the stars. Evaporation of the stars out of the system as a result of the collision was first investigated by L. Spitzer in 1940 [16].

From thermodynamics it is known that evaporation is a process in which molecules having sufficient energy (velocity) to overcome the attraction of neighboring molecules (i.e. fastest) escape beyond the substance (fluid). Wherein, the kinetic energy of the remaining molecules becomes smaller. This leads to cooling of the fluid, and reduction of its weight, and the amount of substance. In a sense, the stars flying away from the cluster can be considered similarly. But there is an important difference. If for the equilibrium systems theory of evaporation can be constructed on the basis of statistical physics and thermodynamics, for systems with the gravity such a theory does not exist. Therefore, numerical calculations are necessary.

We construct the dependence of the fraction of particles "flown" away from the cluster on time for the initial conditions $2K = -U$ (Fig. 3.4). For this case there is evaporation of particles which is characteristic of stellar systems. The largest amount of particles evaporates at the initial stage of the system due to the fact that particles become high speeds during the formation of the core. And during the evolution the particles continue to evaporate so that the fraction increases linearly.

Note that for a system consisting of more than two bodies, particle acquisition of positive total energy, strictly speaking, is not enough to conclude that it will fly to infinity. In a sufficiently dense systems repeated close encounters of bodies are possible as a result of which the total energy may again be negative. However, in the first approximation (as not too tight systems are considered), we assume that it is the particle acquisition of the positive total energy results in its flight to infinity.

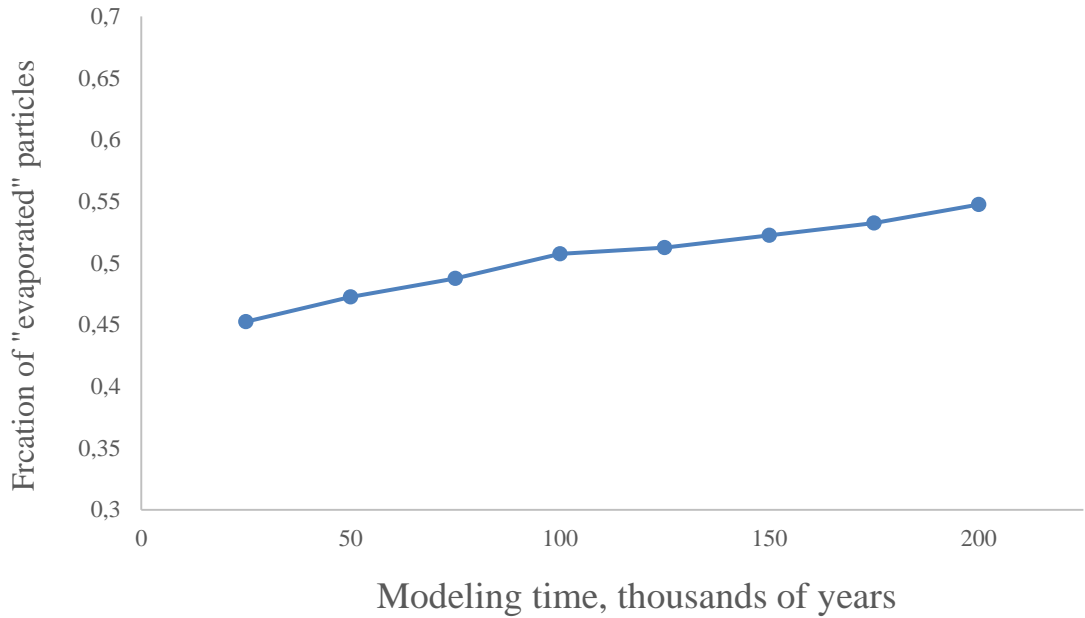

Fig. 3.4 - Dependence of the fraction of evaporated particles from time to time

In order to visualize the characteristic sizes and velocities of the particles, it is possible to introduce the dimensionless variables. In similar problems the following values have been selected as dimensionless variables of time and speed (for example, see [17]):

$$t_0 = \sqrt{\frac{r_m^3}{GM}}, \qquad v_0 = \sqrt{\frac{GM}{r_m}}.$$

If we take half of the cube edge as a characteristic size, we will obtain the characteristic velocity:

$$v_0 = \sqrt{\frac{GM}{r_m}} \approx 8.23 km/s \quad t_0 = \sqrt{\frac{r_m^3}{GM}} \approx 310000 y.$$

In this case during the maximum calculation time with a speed of 10 km/s a particle will fly about

$$l = 6.67 l.y.$$

On the other hand, the average distance between the particles at their uniform initial distribution will be:

$$d \approx 0.7 l.y.$$

These estimates allow us to judge that during the calculations the particles on the average, pass a distance significantly greater than the initial distance between them, but comparable to the size of the original cluster.

It should be noted in this regard that, according to, for example, [15] the characteristic relaxation time of a system with gravity diverges with the increasing number of particles. This leads to the fact that during the characteristic calculation time we can observe, in fact, only the initial stage of such a relaxation.

### 3.2. The distribution on particle coordinates in a regular grid with a ratio of kinetic and potential energy of 2K = -U

Consider now the case with the same initial ratio of kinetic and potential energies *2K = -U*, but the particles are distributed in space not randomly, but are situated in the nodes of a regular grid. As a result of the numerical calculations we construct a velocity distribution function for the 3 time marks - 25, 50, 200 thousand years (Fig. 3.5).

Velocity distribution function over time, as well as in the previous experiment, changes its appearance. At the initial stage of evolution particles have the highest velocities on the same principle - the particles get high velocities because of the aspiration to the center of mass to form the core of the cluster, then the distribution function is gradually shifted to the left in the course of evolution, i.e., there is some relaxation, as a whole velocities decline.

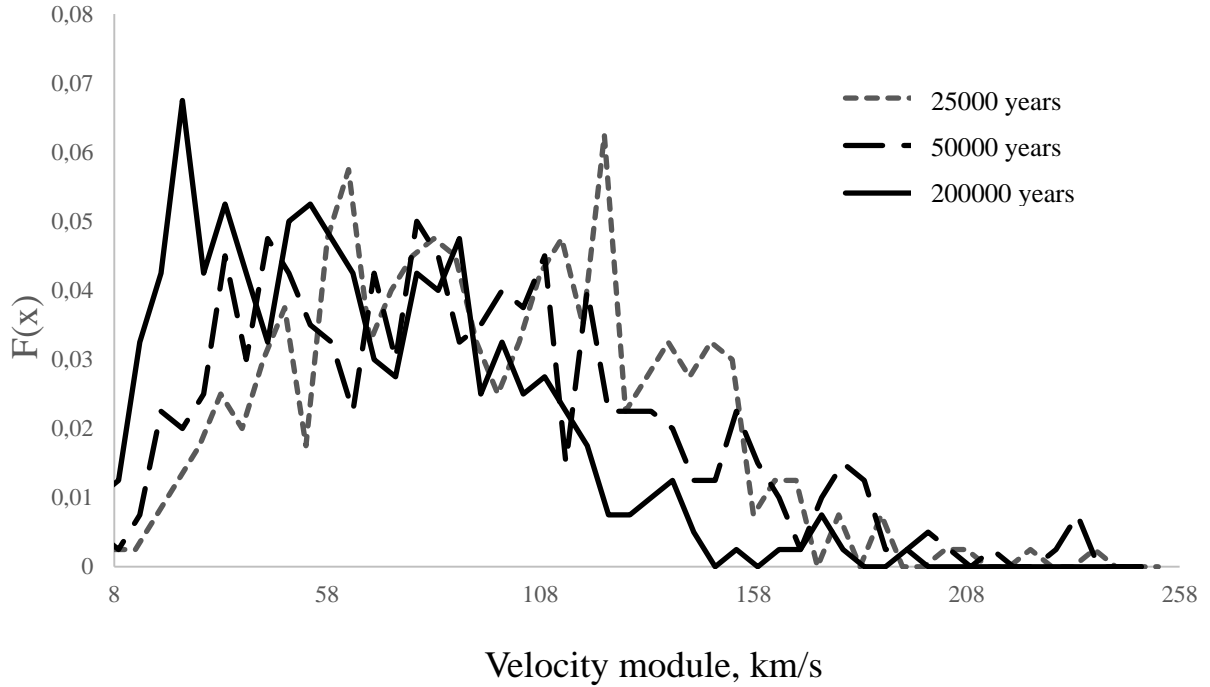

Fig. 3.5 - Distribution of material points in the velocities in the course of evolution. F (x) is the density of probability, normalized value.

Comparing the results of two numerical experiments (with random distribution of the particles on the coordinates and a regular grid), it can be concluded that the second case gives a more uniform distribution of velocities. The distribution function is more "flattened": in the case of random particle distribution peak F(x) reaches almost 0.08, and the bulk of the peaks is in a range from 0.03 to 0.06, in the present case - the peak F(x) reaches a value of only 0,065, and the bulk of the peaks is in a range from 0.02 to 0.05. Velocities in this case are distributed uniformly over the interval.

We construct the dependence of the fraction of evaporated particles on time in this case (see Fig. 3.6).

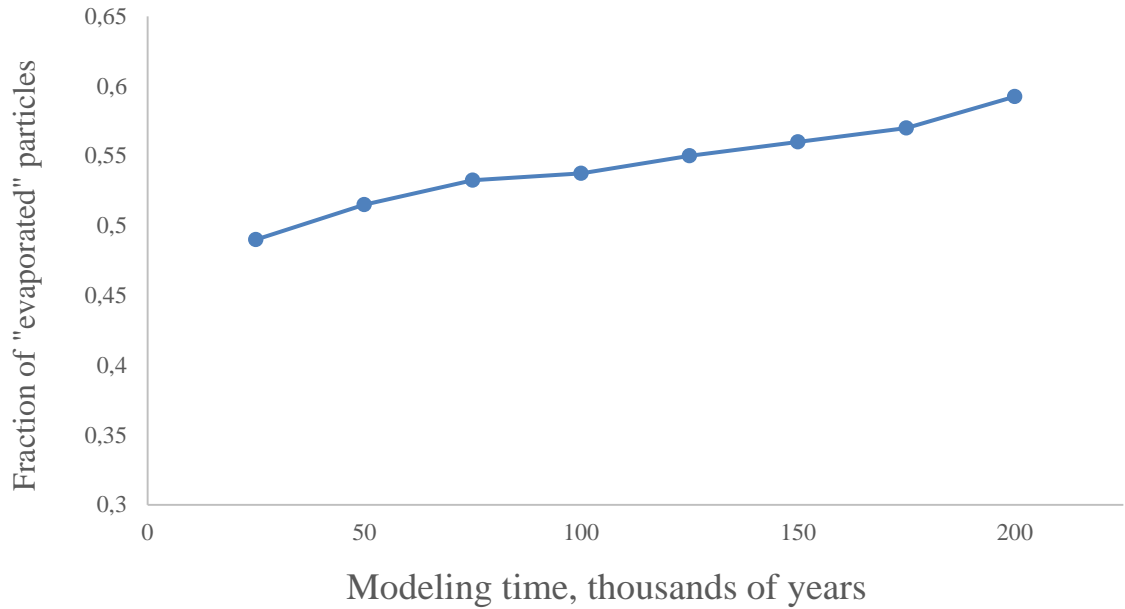

Fig. 3.6 – Fraction of evaporated particles in the case of a ratio of kinetic and potential energies $2K = -U$ in a regular grid.

There is evaporation of particles, characteristic of stellar systems. Significant growth of particles evaporation occurs in the initial stage due to the fact that the particles acquire high velocities during the formation of the core. The particles evaporate from the system more evenly.

When comparing the evolution of the two systems (with the ratio of the kinetic and potential energy $2K = -U$ with a random distribution of particles in a regular grid with the same velocity distribution), one can conclude that the distribution in a regular grid allowed the particles to gain higher velocities (see. Fig. 3.1 and 3.5), and this, in turn, caused a great expansion of clusters of particles (see Fig. 3.4 and Fig. 3.6).

The obtained results can be compared qualitatively with the results derived by other authors earlier. In [17-19] the evolution of gravitating systems was considered, in particular, in the case of fulfillment of the virial theorem $2K = -U$. Such conditions allow to avoid the effect of particle correlations (collisions). In particular, it has been obtained that at low virial numbers (i.e. at low kinetic energy of the particles), the initial symmetrical distribution of particles in space is violated and there arises a spindle-shaped structure. In the next section we consider the ratio between kinetic and potential energy equal to *8K = -U*. In our case, a significant violation of symmetry is not observed. Perhaps this is due to a somewhat different virial number, or due to the fact that the number of particles is relatively small (compared with [17]). This could lead to the fact that a relatively small symmetry breaking is difficult to observe on the background of random initial distribution of particles.

We give an example of dependence of total energy of the particles randomly selected at the initial moment of time in the considered cluster (Fig. 3.7, 3.8).

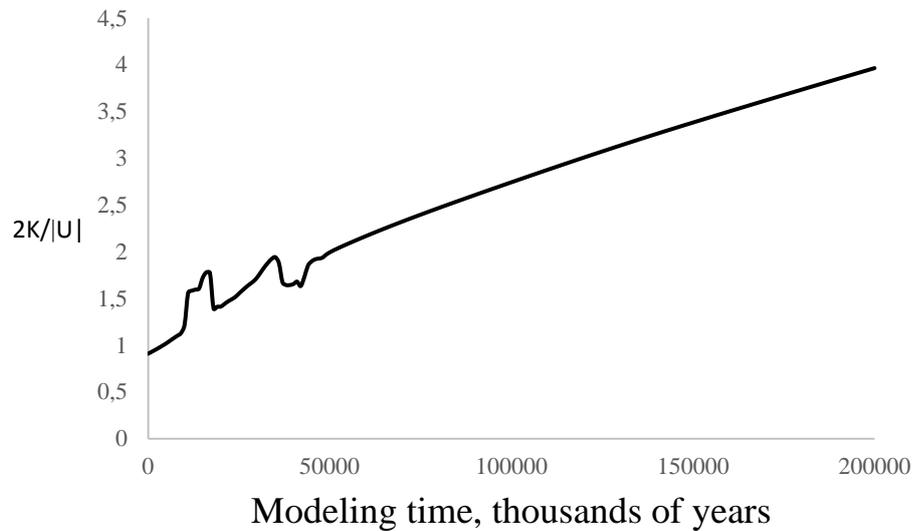

Fig. 3.7. Dependence of *2K/U* ratio for an evaporating particle on time

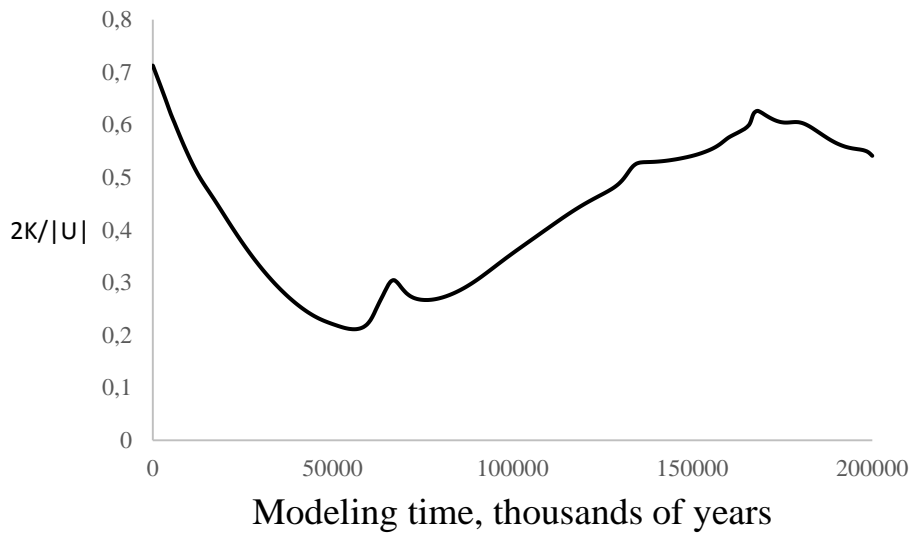

Figure 3.8. Dependence of *2K/U* ratio for a non-evaporating particle on time

We note that in the framework of the model considered the fulfillment of the virial theorem takes place for a system as a whole. However, for a given initial condition different particles have different potential energy. The most significant differences are observed for particles located in the vertices of a cube and inside it. This leads to the fact that the ratio of $2K/U$ for different particles is different - for some particles it is less than one, for others – more than one.

**3.3 The distribution of particle coordinates with the ratio of potential and kinetic energies of $8K = -U$**

To study the evolution of gravitating system the initial distribution of the particles on the coordinates was considered, the same as for the experiment with the ratio of the kinetic and potential energy $2K = -U$ with a random distribution of the particles. The cluster has a linear size 5 light years. The kinetic and potential energy of the system relate as $8K = -U$. The velocity distribution coincides with that of the previous experiment. The components of the velocity of the particles are equal in magnitude and equiprobable in direction. Module of velocity is 4 km / s.

According to the results of modeling, velocity distribution function for 3 time marks was constructed - 25, 50, 200 thousand years (Fig. 3.9).

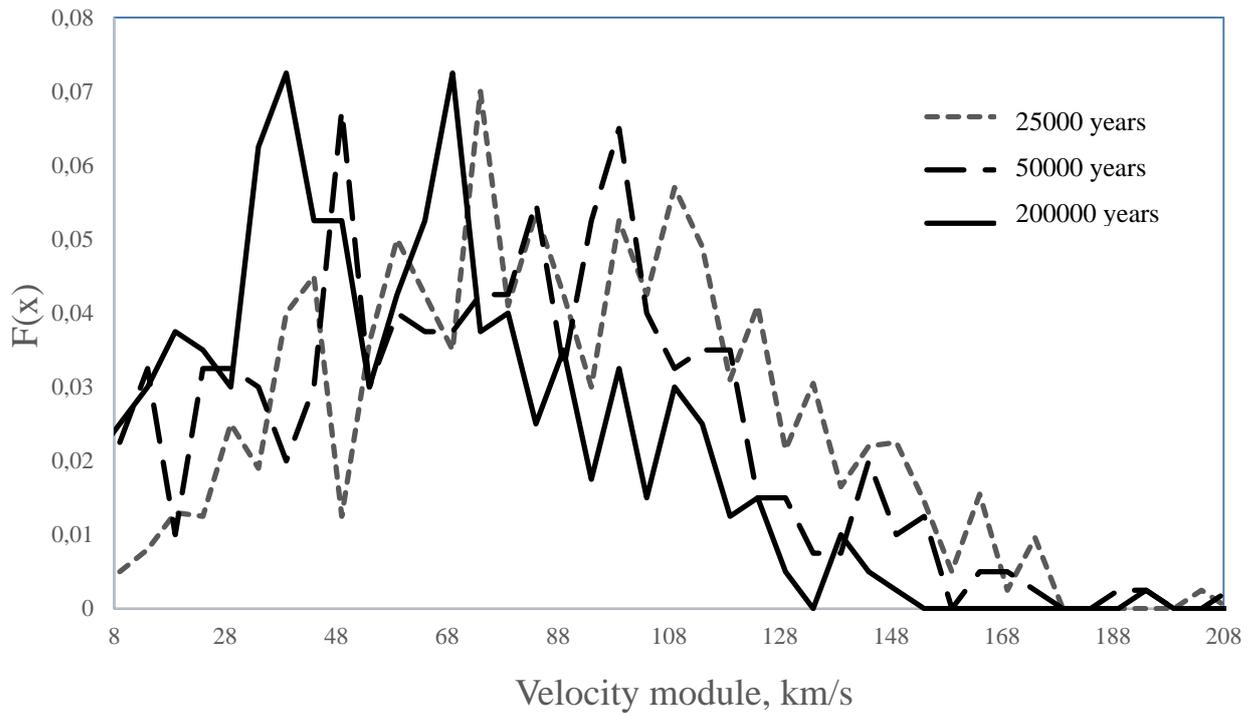

Fig. 3.9 – Velocity distribution function. F(x) is the density of probability, normalized value.

The maximum of distribution function during the evolution shifts to the left (the velocities of particles decrease). The process of change and displacement/shifting of distribution function is identical to the previous experiments.

For a more detailed analysis to see how the other initial conditions, namely the different initial velocities of the particles affect the distribution function, we compare the distribution function for the ratio of energy *2K = -U* and *8K = -U* (Fig. 3.10).

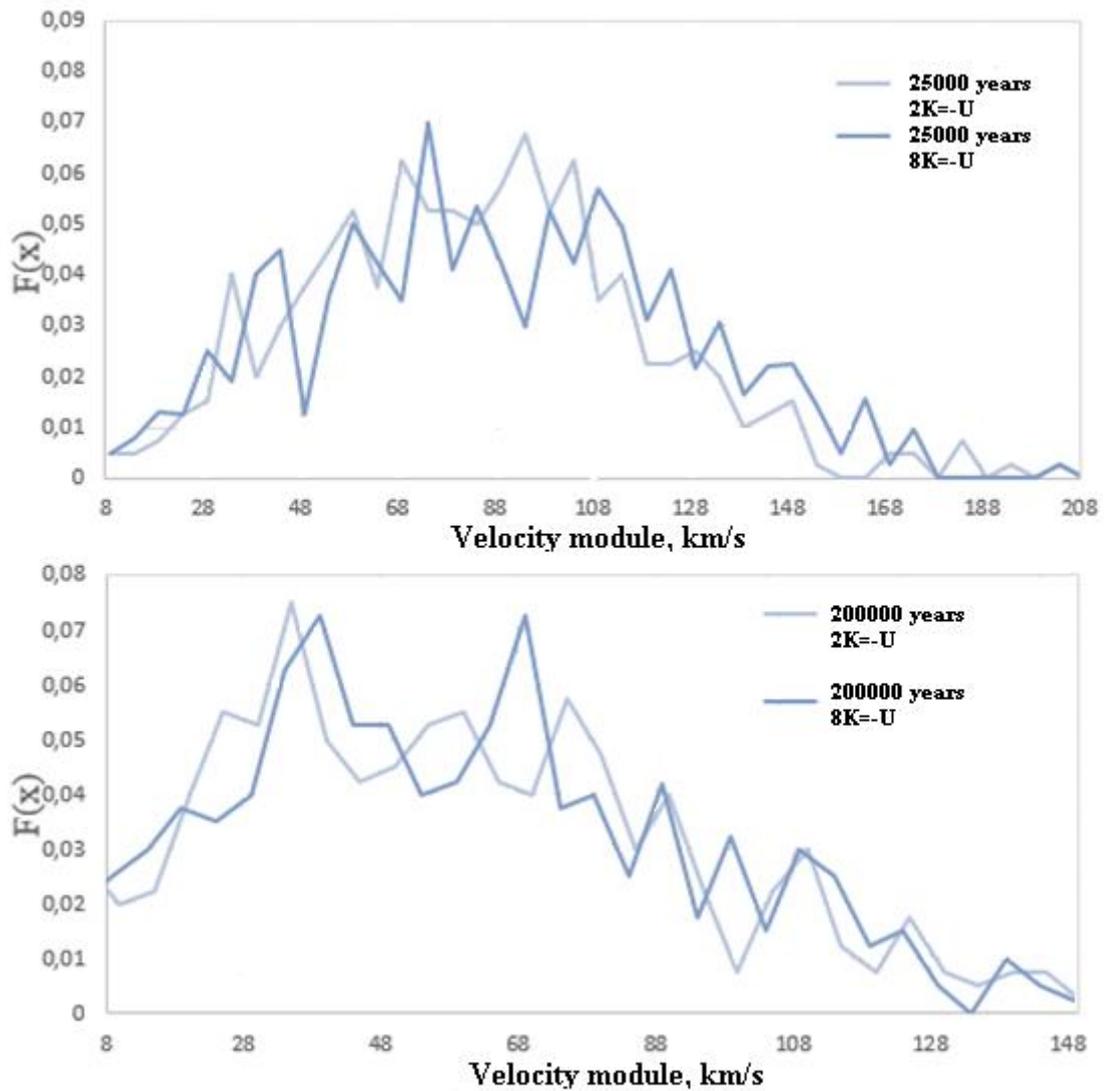

Fig. 3.10 - Comparison of the distribution functions for the two cases. F(x) is the density of probability, normalized value.

From the results of numerical experiments it can be concluded that the distribution functions for different kinetic energies are very close to each other (taking into account fluctuations caused by random distribution of particles at the initial moment time). On the other hand, we can follow the evolution of the distribution function at sufficiently long times, which results in the fact that fraction of particles with low velocities becomes smaller.

Spatial distribution of particles with different initial kinetic energy is shown on Fig. 3.11.

**25000 years**

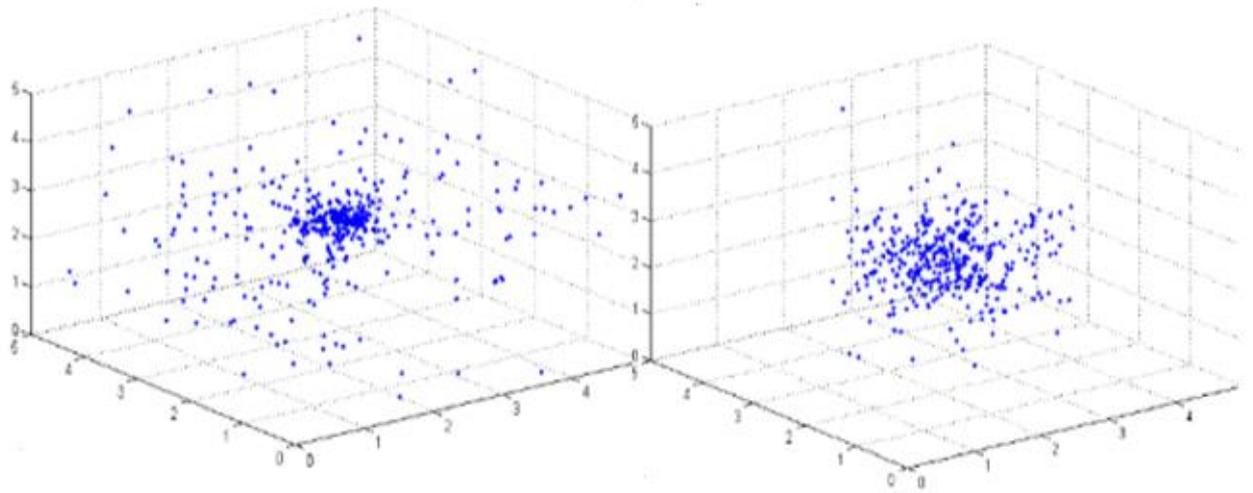

**2K=-U**  **8K=-U**

Fig. 3.11 –The formation of the core of the particles for cases with a ratio of kinetic and potential energy *2K = -U* and *8K = -U*.

The figure 3.11 shows that with the initial conditions 2K = -U at the moment the nucleus had time to be formed due to greater initial velocities of the particles, and in the case of the initial conditions *8K = -U* by this time the particles with lesser initial velocities only begin to move to the center of the cluster.

Let us compare the dependence of the fraction of particles evaporated from a cluster on time for cases with a ratio of energy *2K = -U* and *8K = -U* (Fig. 3.12).

According to the results of numerical calculations (Fig. 3.12) it can be concluded that the number of particles evaporated at an early stage in both cases is almost the same, then this amount increases linearly, and for a case of *2K = -U* evaporation occurs more actively.

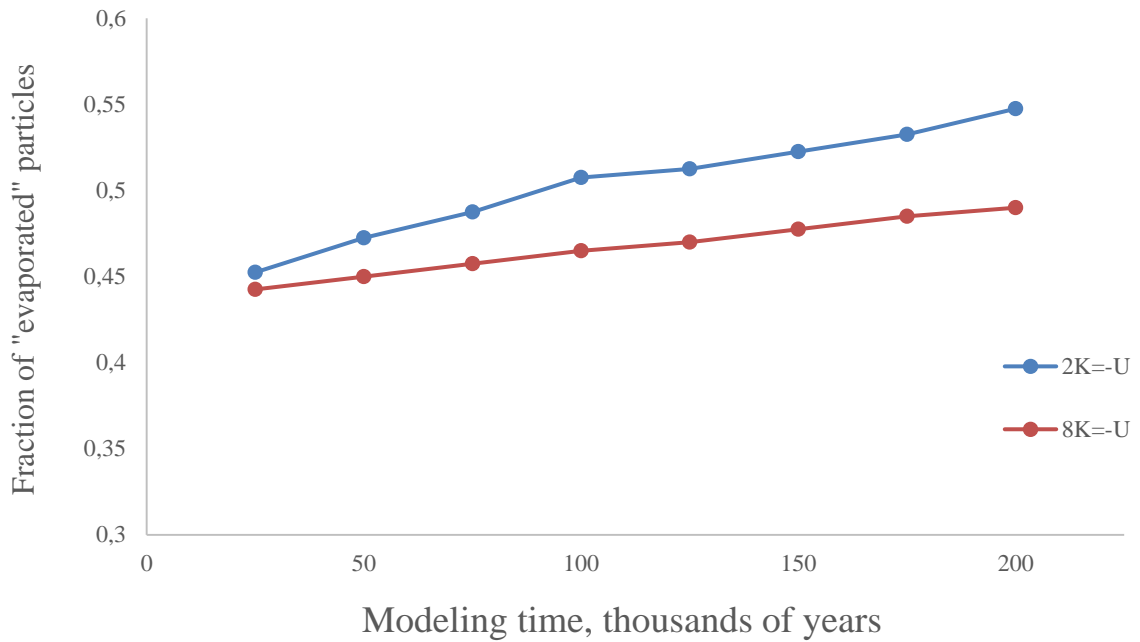

Fig. 3.12 - Fraction of evaporated particles for cases with a ratio of kinetic and potential energies *2K = -U* and *8K = -U*.

## 3.4. The distribution of particles on coordinates with the ratio of potential and kinetic energies of 5U = -K

To study the evolution of gravitating system the initial distribution of particle coordinates was taken , the same as for the random distribution with ratio of the kinetic and potential energies *2K = -U*. The cluster has a linear size 5 light years. The kinetic and potential energy of the system relate as *5U=-K*. The components of the velocity of the particles are equal and equiprobable in direction. Module of velocity is 31 km / s.

Velocity distribution function for 3 time marks was constructed - 25, 50, 200 thousand years (Fig. 3.13).

The distribution function for the given initial conditions looks narrower when compared with other initial conditions. At the initial stage of evolution particles have the greatest velocities in comparison with the other experiments, which is the result of the greatest initial kinetic energy. The time dependence of the velocity distribution function of particles is shown on Fig. 3.13.

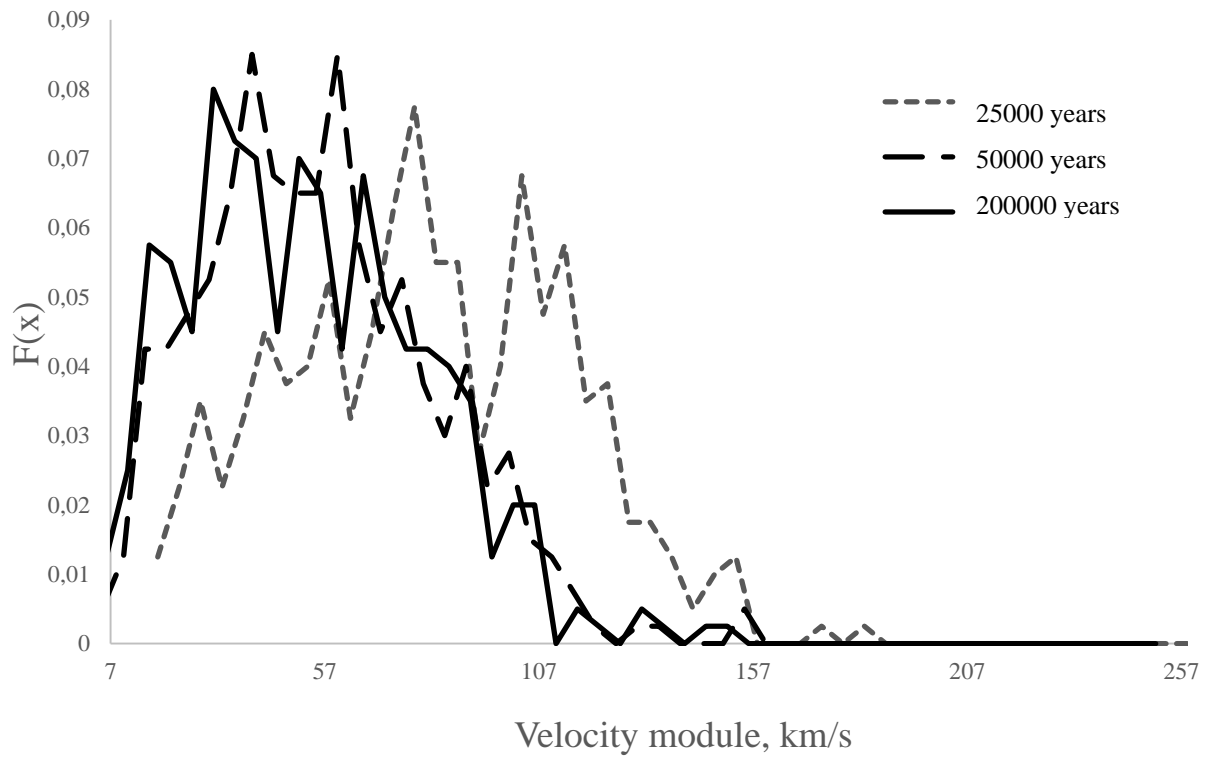

Fig. 3.13- The velocity distribution of material points in dependence on time for the case *5U=-K*. F(x) is the density of probability, normalized value.

Let us compare the dependence of the fraction of evaporated particles in all cases (Fig. 3.14). We can see that the greater initial kinetic energy of particles is, the larger is the fraction of evaporated particles. The result for the system with distribution of particles in a regular grid is less obvious, because for this case kinetic energy coincides with energy for random distribution of particles.

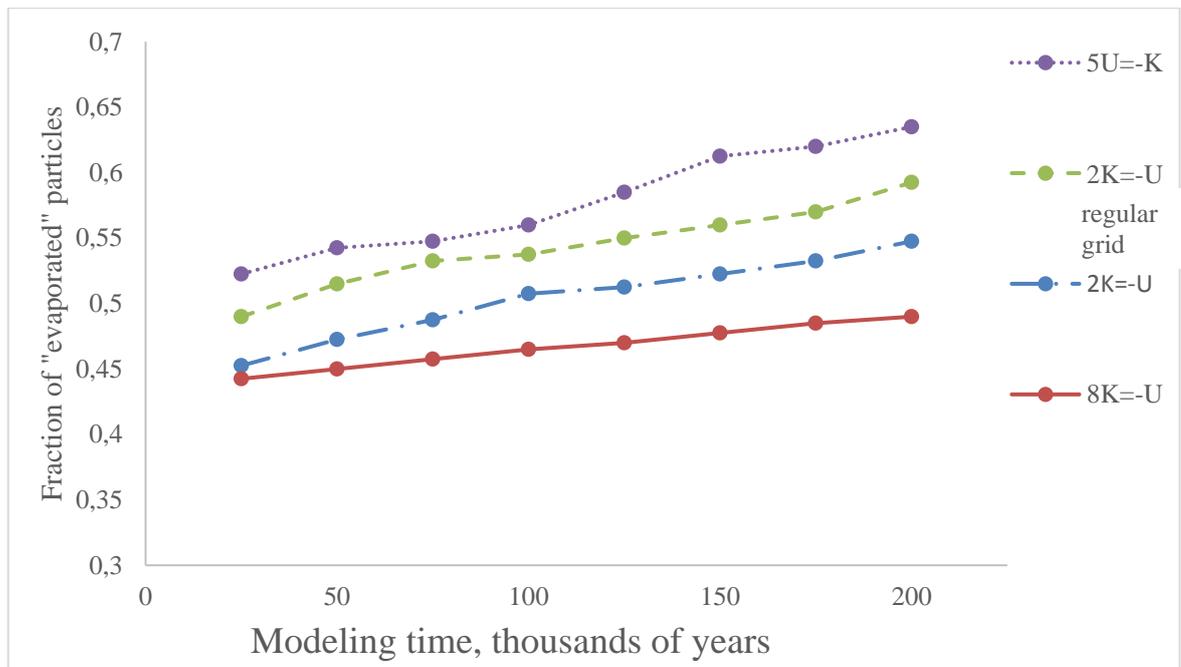

Fig. 3.14 - Fraction of evaporated particles for all the considered cases.

During 200,000 years the particles on average have time to fly a distance equal to the 6-8 initial cluster size (30-40 light-years) or about 6-8 periods of revolution around the core of the cluster.

# 4. Investigation of the diffusion of gravitating systems with different initial conditions

The term "diffusion" in relation to the stellar dynamics can be applied only approximately, because, strictly speaking, it can only be introduced for the systems with local equilibrium. In fact, in this case, when local equilibrium is not presented, we can say about the mixing of particles.

The study of diffusion was conducted for all of the cases examined in the previous chapter. For this purpose, the space was divided by plane YZ (X = 2,5), passing through the center of the cluster, into two parts, and particles are assumed as two different classes (Fig. 4.1). All other properties of particles of both classes are the same.

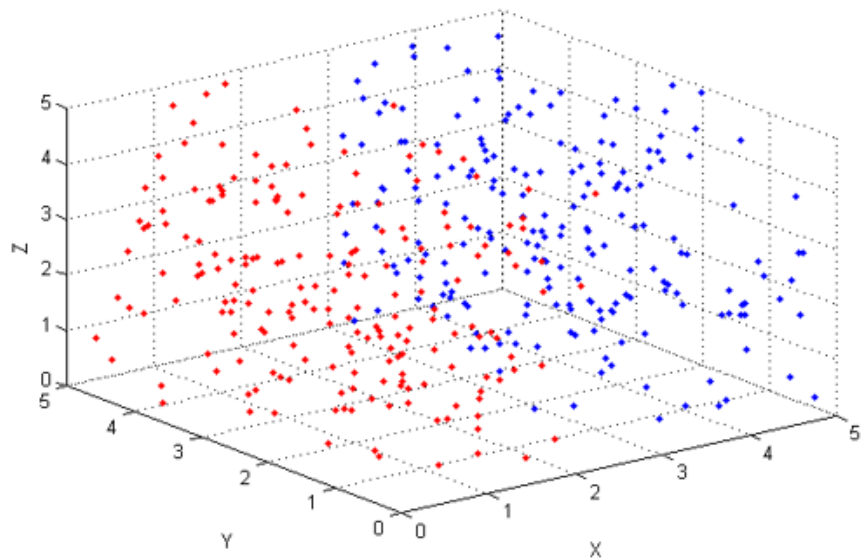

Fig. 4.1 - 3D cluster has a size of 5 l.y. and is divided by YZ plane for the study of diffusion

System status at time point of 25 thousand years for the initial conditions $2K = -U$ is shown in Fig. 4.2.

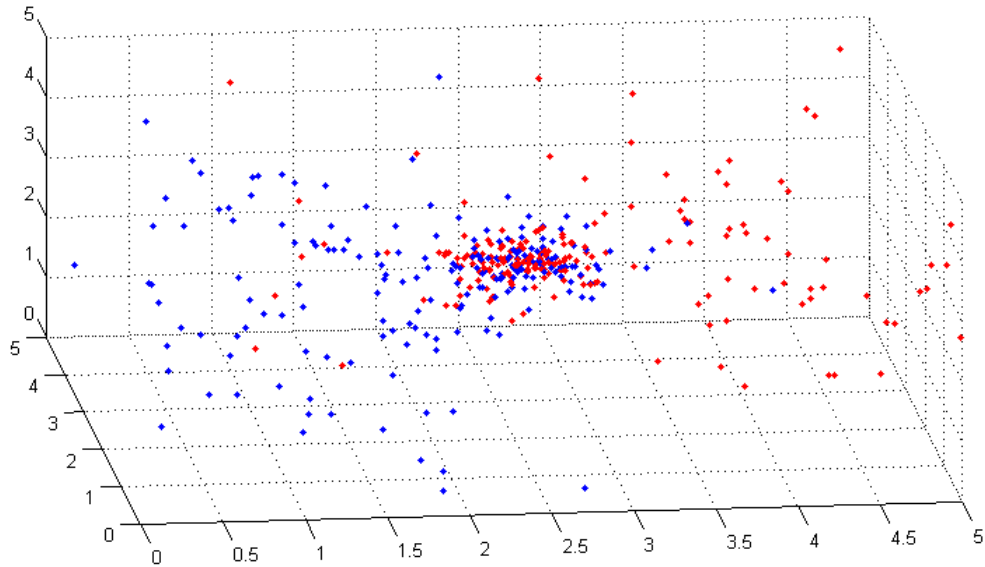

Fig. 4.2 - Status of the cluster at time point of 25 thousand years for the initial conditions *2K = -U*

According to the graph, shown in Figure 4.3, it is clear that the first time point of 25 thousand years, much of both types of particles is moving in the opposite (from the one in which they originally were) part of the cluster. At the beginning of evolution the particles from cluster edges aspire to its the opposite part. That is why during the first stage the largest number of particles goes to infinity (Fig. 3.4) as a large fraction of the particles purchases significant velocities.

During mutual moving of particles in opposite directions, the fraction of particles flies farther, and some fly out from the cluster, and the other part stays in the center of the cluster because of the close mutual flights of particles.

In the early stages as a result of random distribution the probability of the fact that two particles will be close enough is smaller than in much later stage because of mutual attraction of particles.

The core of a large mass begins to form, since most of the particles are concentrated in some finite central part of the cluster, which is displayed in Fig. 4.2. From the core, which contains a large number of particles, over time the particles continue to fly because of the mutual influence on each other. In Fig. 4.2 we can see that the core formed slightly to the right from center of the cluster, which confirms the prevalence of both classes precisely in the right side of the cluster.

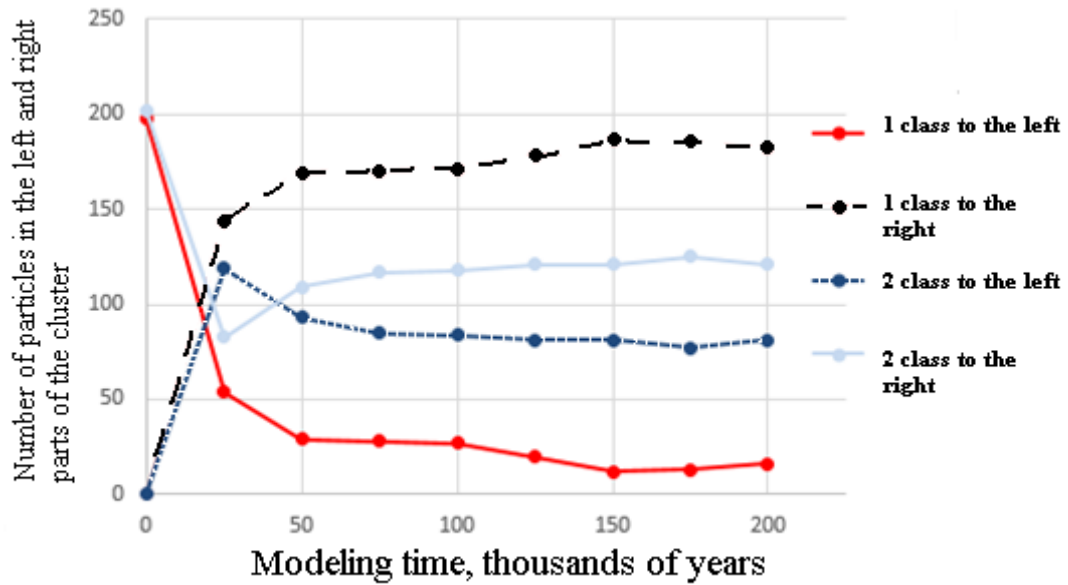

Fig. 4.3 - Dependence of the number of particles of 2 classes in t left and right sides of the cluster on the time for the case of $2K = -U$

For the case of $8K = -U$ the distribution of particles classes is more uniform (Figure 4.4). It indicates that the core is in the central part of the cluster. There is a good spatial mixing (Figure 4.5), but over time the full mixing does not occur.

The dependence of the number of particles of 2 classes in the left and right sides on time for cases $5U = -K$ and $2K = -U$ in a regular grid are similar to each other (Fig. 4.6 and 4.7).

Thus, the alignment of number of the two classes of particles in the left and right sides of the cluster does not occur for all initial conditions investigated in the present work. This can be explained by the fact that for the systems with long-range potential the characteristic time of mixing is much larger than characteristic time of particle moving along its orbit.

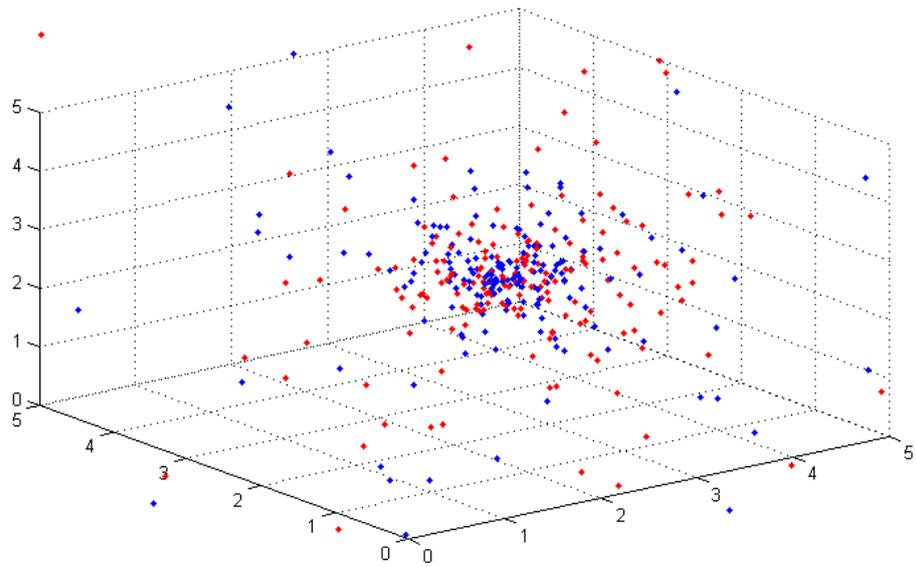

Fig. 4.4 – Cluster status of at time point of 175 thousand years for the initial conditions *8K = -U*

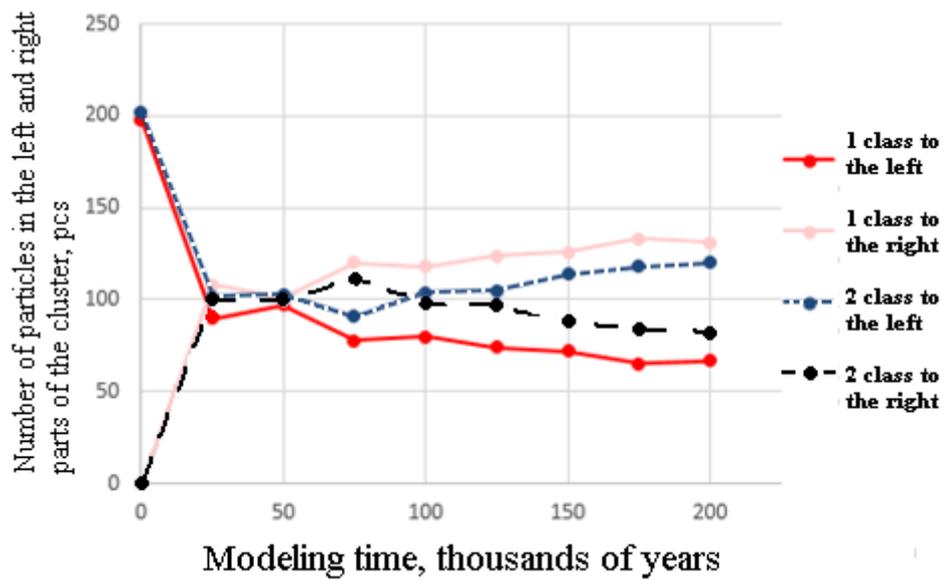

Fig. 4.5 - Dependence of the number of particles of 2 classes in left and right sides of the cluster on the time for the case *8K = -U*

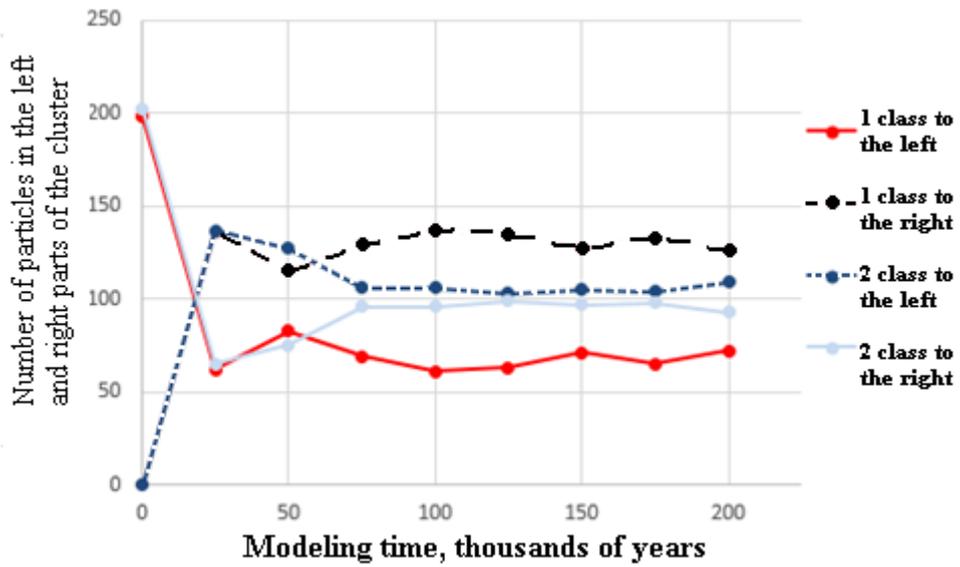

Fig. 4.6 - Dependence of the number of particles of 2 classes in left and right sides of the cluster on the time for the case $5U=-K$

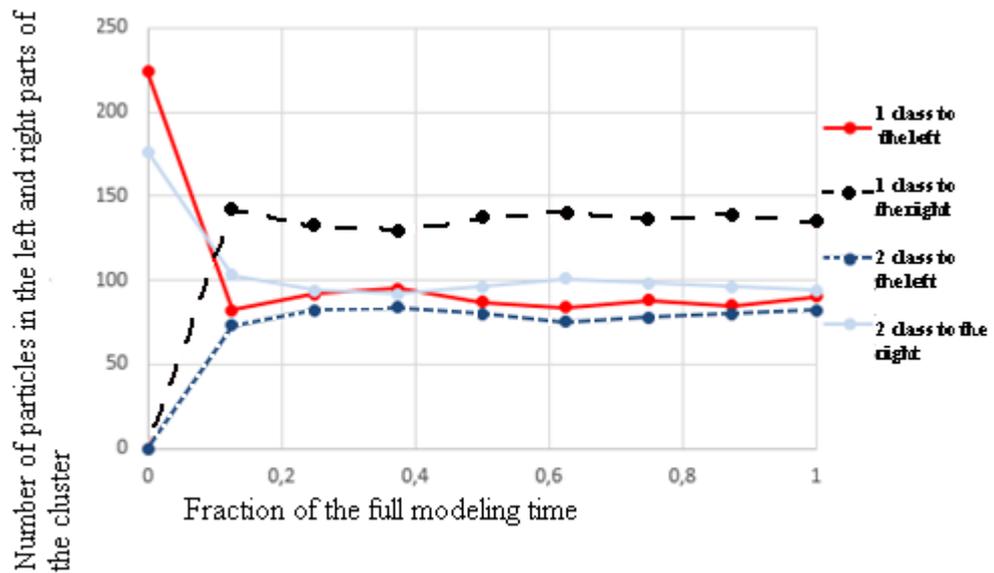

Fig. 4.7 - Dependence of the number of particles of 2 classes in left and right sides of the cluster on the time for the case $2K=-U$ in a regular grid

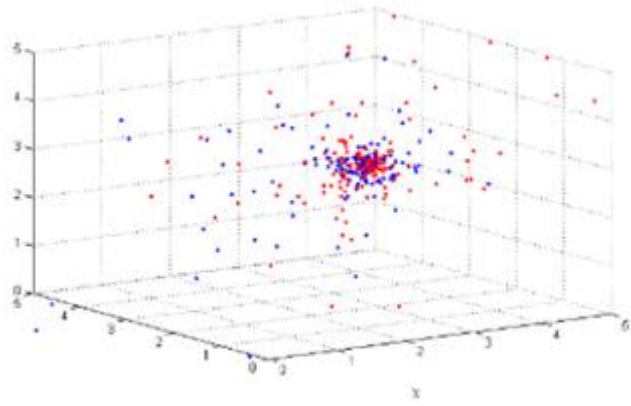 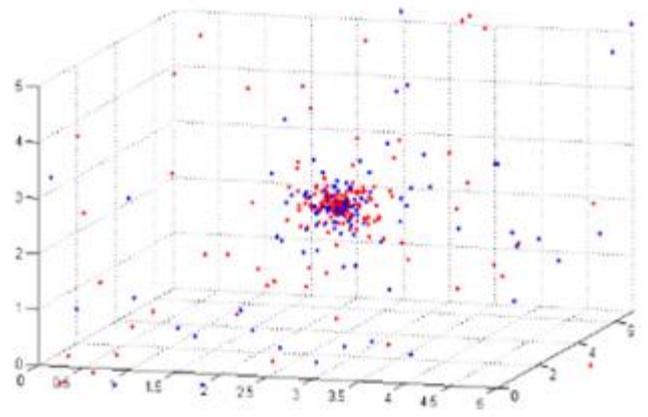

**5U=-K**  **2K=-U regular grid**

Fig. 4.8 - Cluster status of the at time point of 50 thousand years for the initial conditions $5U = -K$ and $2K = -U$ in a regular grid

## 5. Maximum Lyapunov exponent of a gravitating system with different initial conditions

Since it is the correlation between the particles that causes them to evaporate from the system, it is important to trace the relationship between the characteristics of chaos in the system (for example, the characteristic time of the phase divergence of the trajectories) and the parameters of the dynamic system.

As a characteristic of this chaotization a maximum Lyapunov exponent can act:

$$h = \lim_{\substack{d(0) \to 0 \\ t \to \infty}} \frac{1}{t} \ln \frac{d(t)}{d(0)}, \quad (5.1)$$

where $d(0)$ - the initial distance between the points in phase space.

The reciprocal

$$\tau = \frac{1}{\lambda}, \quad (5.2)$$

is the characteristic time of divergence of trajectories. That is, we study the evolution of close points and find to what extent their trajectories diverge over time. Measure of divergence of the trajectories is the distance between the points $d(t)$.

Using the maximum Lyapunov exponent, we can determine what is the investigated mode - chaotic or regular. In particular, if the system dynamics is periodic or quasi-periodic, than over time the distance $d(t)$ does not increase and the maximum Lyapunov exponent is zero ($\lambda = 0$). In the case of chaotic motion $\lambda > 0$.

For the four initial conditions maximum Lyapunov index was calculated (Fig. 5.1). For the planets of the solar system the value of $\lambda$ (5.1) is, for example, of the order of $10^{-5}$-$10^{-6}$ [20]. Since the resonances and the associated dynamic chaos are responsible for the evaporation of the particles, the average particle evaporation time should be about the average time of divergence of the trajectories.

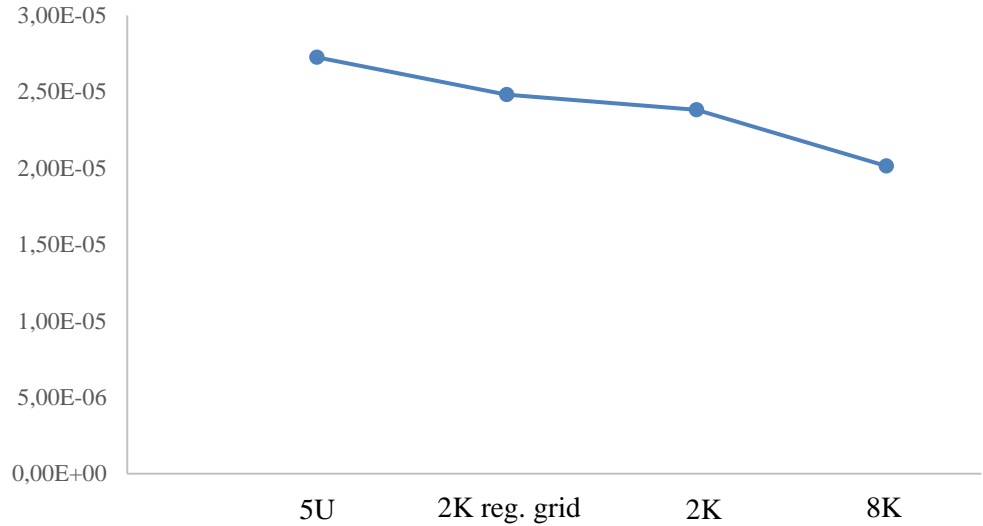

Fig.5.1 - The maximum Lyapunov exponent of systems for four different cases of initial conditions

Its reciprocal is estimated - characteristic time of divergence of trajectories varies from 40 to 50 thousand years. In dimensionless quantities this time is approximately 1/7 of $t_0$.

This value is of the same order with the characteristic time, during which the fraction of evaporated particles changes substantially (Fig. 3.14). The system at the initial conditions with a higher kinetic energy $5U = -K$ has a more chaotic mode as the maximum Lyapunov exponent in this case is higher than for systems with initial conditions $2K = -U$ and $8K = -U$. This may cause more intense evaporation of particles in such conditions (see Fig. 3.14).

# Conclusion

On the basis of numerical modeling the evolution of systems with gravity at different initial ratio between the kinetic and potential energy of the particles is studied. As a result of simulation the time dependence of the velocity distribution function for the system, consisting of material points, during the evolution is obtained. The dependence of the fraction of evaporated particles from the cluster on time for different initial conditions is obtained. The fraction of these particles depends on the initial conditions so that for higher initial velocities, it is about 0.65, while for the lowest ones it is 0.45.

Diffusion within gravitating system consisting of two kinds of particles is investigated. It is shown that the characteristic mixing time is much greater than the characteristic time of particle motion along its orbit.

The maximum Lyapunov exponent of systems with different initial conditions is calculated. The values obtained were approximately $10^{-5}$. At the same time, the characteristic time of divergence of particle trajectories - 40-50 thousand years. The obtained results can be used to understand the origin and evolution of 3D gravitating systems.

This work was partially supported by Act 211 Government of the Russian Federation, agreement № 02.A03.21.0006 and by the Russian Foundation for Basic Researched (*RFBR*) under Grant No. 16-31-00274.